\documentclass[twocolumn,aps,prb,floatfix,showpacs,showkeys,amssymb,superscriptaddress]{revtex4}
\usepackage{graphicx,epsf,dcolumn,subfigure}
\begin{document}

\def\baselinestretch{1.0}

\begin{titlepage}
\date{\today}

\title{Electronic structure and optical properties of Zn$X$ ($X$=O, S, Se, Te)}

\author{S. Zh. Karazhanov}

\affiliation{Centre for Material Science and Nanotechnology,
Department of Chemistry, University of Oslo, PO Box 1033 Blindern,
N-0315 Oslo, Norway}

\affiliation{Physical-Technical Institute, 2B Mavlyanov St.,
Tashkent, 700084, Uzbekistan}

\author{P. Ravindran}

\affiliation{Centre for Material Science and Nanotechnology,
Department of Chemistry, University of Oslo, PO Box 1033 Blindern,
N-0315 Oslo, Norway}

\email[Corresponding author:]{ponniah.ravindran@kjemi.uio.no}

\author{A. Kjekshus}

\affiliation{Centre for Material Science and Nanotechnology,
Department of Chemistry, University of Oslo, PO Box 1033 Blindern,
N-0315 Oslo, Norway}

\author{H. Fjellv{\aa}g}
\affiliation{Centre for Material Science and Nanotechnology,
Department of Chemistry, University of Oslo, PO Box 1033 Blindern,
N-0315 Oslo, Norway}

\author{B. G. Svensson}

\affiliation{Department of Physics, University of Oslo, PO Box
1048 Blindern, N-0316 Oslo, Norway}

\begin{abstract}

Electronic band structure and optical properties of zinc
monochalcogenides with zinc-blende- and wurtzite-type structures
were studied using the \textit{ab initio} density functional
method within the LDA, GGA, and LDA+$U$ approaches. Calculations
of the optical spectra have been performed for the energy range
0--20~eV, with and without including spin-orbit coupling.
Reflectivity, absorption and extinction coefficients, and
refractive index have been computed from the imaginary part of the
dielectric function using the Kramers--Kronig transformations. A
rigid shift of the calculated optical spectra is found to provide
a good first approximation to reproduce experimental observations
for almost all the zinc monochalcogenide phases considered.  By
inspection of the calculated and experimentally determined
band-gap values for the zinc monochalcogenide series, the band gap
of ZnO with zinc-blende structure has been estimated.

\end{abstract}



\maketitle

\end{titlepage}

\normalsize
\section{\label{intro} Introduction}

The zinc monochalcogenides (Zn$X$; $X$ = O, S, Se, and Te) are the
prototype II--VI semiconductors. These compounds are reported to
crystallize in the zinc-blende(-z) and wurtzite (w) type
structures. The Zn$X$-z phases are optically isotropic, while the
Zn$X$-w phases are anisotropic with $c$ as the polar axis. Zn$X$
phases are primary candidate for optical device technology such as
visual displays, high-density optical memories, transparent
conductors, solid-state laser devices, photodetectors, solar cells
etc. So, knowledge about optical properties of these materials is
especially important in the design and analysis of Zn$X$-based
optoelectronic devices.

\begin{figure*}
\centering
\includegraphics[scale=0.75]{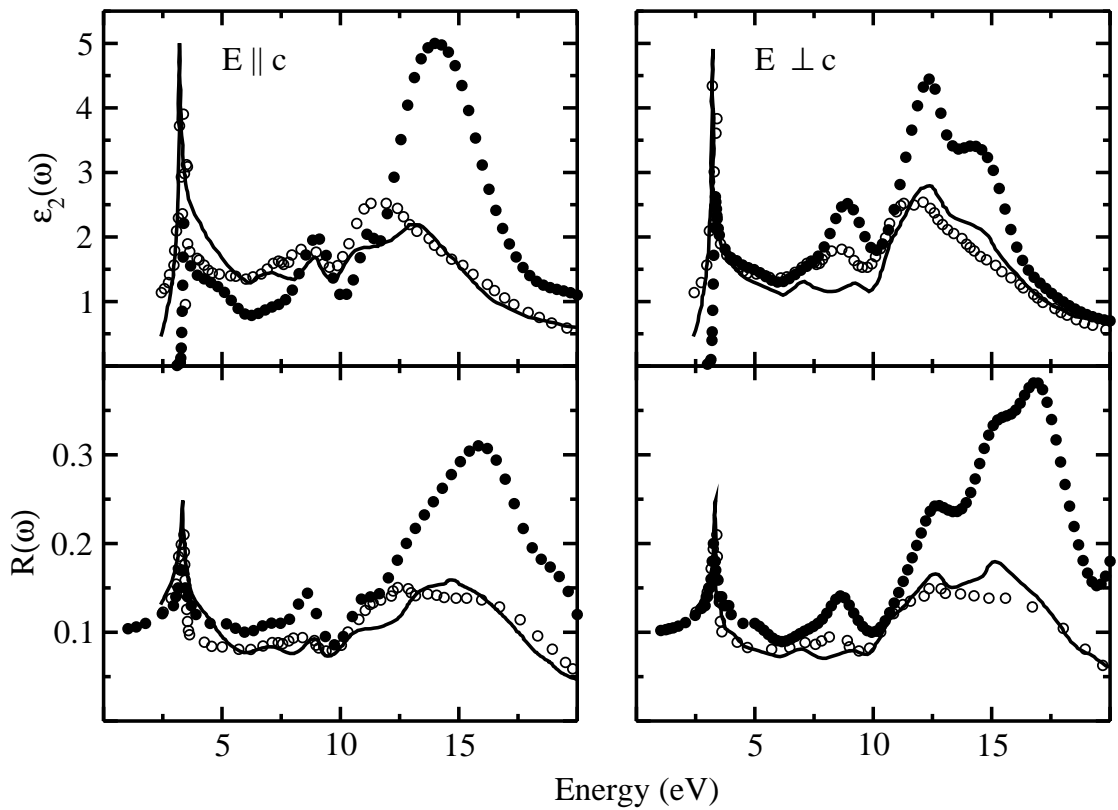}
\caption{Reflectivity spectra $R(\omega)$ for ZnO-w determined
experimentally at 300~K in Refs.~\onlinecite{A99,F73} (solid
circles), Ref.~\onlinecite{KNPSW71} (open circles), and
Ref.~\onlinecite{HAP70} (solid lines) along with imaginary part of
the dielectric response function ($\epsilon_2(\omega)$) calculated
using the Kramers-Kronig relation. The results of
Ref.~\onlinecite{HAP70} (open circles) are used for both $E
\parallel c$ and $E \perp c$, because no polarized incident light was
used in the experiments.} \label{exp_ZnO}
\end{figure*}

Optical parameters for some of the Zn$X$ phases have widely been
studied experimentally in the past. Detailed information on this
subject is available for
ZnO-w,\cite{YA97,PSAYIIH00,PSAYMI01,WODC98,MKSON99,SC98,HAP70,KNPSW71,F73}
ZnS-w,\cite{F73} ZnS-z,\cite{F73,CH70,CH65} ZnSe-z,\cite{F73,CH70}
and ZnTe-z,\cite{F73,CH70,MKMCA05,CG63} see the systematized
survey in Ref.~\onlinecite{A99}. However, there are no
experimental data on optical properties of ZnSe-w, ZnTe-w, and
ZnO-z. Furthermore, there is lack of consistency between some of
the experimental values for the optical spectra. This is
demonstrated in Fig.~~\ref{exp_ZnO}, which displays reflectivity
spectra for ZnO-w measured at T=300~K by three different groups.
Dielectric response functions were calculated using the
Kramers-Kronig relation. As is seen in Fig.~\ref{exp_ZnO},
intensity of the imaginary part of the dielectric function
($\epsilon_2$) and reflectivity ($R$) corresponding to the
fundamental absorption edge of ZnO-w are higher\cite{KNPSW71} than
those at the energy range 10--15~eV, while in
Ref.~\onlinecite{A99} it is vice verse. The optical spectra in
Fig.~\ref{exp_ZnO} measured using the linearly polarized incident
light for electric field ($E$) parallel $(\parallel)$ and
perpendicular $(\perp)$ to the $c$-axes are somehow close to those
of Ref.~\onlinecite{HAP70} using non-polarized incident light.

Using the experimental reflectivity data, a full set of optical
spectra for ZnO have been calculated\cite{SS03} for the wide
energy range 0--26~eV. Density-functional theory (DFT)\cite{HK64}
in the local-density approximation (LDA)\cite{KS65} has also been
used to calculate optical spectra for ZnO-w\cite{XC93} and
ZnS-w\cite{XC93} by linear combination of atomic orbitals, and for
ZnS-z\cite{WK812} and ZnSe-z\cite{WK812} by self-consistent linear
combination of Gaussian orbitals. The optical spectra of ZnO
(including excitons) has been investigated\cite{LC06} by solving
the Bethe--Salpeter equation. Band-structure studies have been
performed by linearized augmented plane-wave method plus local
orbitals (LAPW+LO) within the generalized gradient and LDA with
the multiorbital mean-field Hubbard potential (LDA+$U$)
approximations. The latter approximation is found to correct not
only the energy location of the Zn-3$d$ electrons and associated
band parameters (see also
Refs.~\onlinecite{KRGKFS051}~and~\onlinecite{KRGKFS052}), but also
to improve the optical response. Despite the shortcoming of DFT in
relation to underestimation of band gaps, the locations of the
major peaks in the calculated energy dependence of the optical
spectra are found to be in good agreement with experimental data.

It should be noted that the error in calculation of the band gap
by DFT within LDA and generalized-gradient approximation (GGA) is
more severe in semiconductors with strong Coulomb correlation
effects than in other
solids.\cite{ASKCS93,DBSHS98,BABH00,KRGKFS051,KRGKFS052} This is
due to the mean-field character of the Kohn--Sham equations and
the poor description of the strong Coulomb correlation and
exchange interaction between electrons in narrow $d$~bands (viz.
the potential $U$). Not only the band gap ($E_g$), but also the
crystal-field (CF) and spin-orbit (SO) splitting energies
($\Delta_{\rm{CF}}$ and $\Delta_{\rm{SO}}$), the order of states
at the top of the valence band (VB), the location of the Zn-3$d$
band and its width, and the band dispersion are
found\cite{LRLSM02,LYVWCC96,KRGKFS051,KRGKFS052} to be incorrect
for ZnO-w by the \textit{ab initio} full potential (FP) and
atomic-sphere-approximation (ASA) linear muffin-tin orbital (LMTO)
methods within the pure LDA,\cite{LRLSM02,LYVWCC96} and by the
projector--augmented wave (PAW) method within LDA and
GGA.\cite{KRGKFS051,KRGKFS052} These findings were
ascribed\cite{KRGKFS051,KRGKFS052} to strong Coulomb correlation
effects. DFT calculations within  LDA plus self-interaction
correction (LDA+$SIC$) and LDA+$U$ are
found\cite{LRLSM02,KRGKFS051,KRGKFS052} to rectify the errors
related to $\Delta_{\rm{CF}}$ and $\Delta_{\rm{SO}}$, order of
states at the top VB, width and location of the Zn 3$d$ band, as
well as effective masses. In other semiconductors, in which the
Coulomb correlation is not sufficiently strong, the
$\Delta_{\rm{CF}}$ and $\Delta_{\rm{SO}}$ values derived from DFT
calculations within LDA are found to be quite accurate. This was
demonstrated for diamond-like group IV, z-type group III--V,
II--VI, and I--VII semiconductors,\cite{CW04} w-type AlN, GaN, and
InN~\cite{WZ96} using the LAPW and VASP-PAW, the w-type CdS and
CdSe,~\cite{LYVWCC96} z-type ZnSe, CdTe, HgTe,\cite{WCC95} using
the \textit{ab initio} LMTO-ASA, z- and w-type ZnSe and
ZnTe\cite{KRGKFS051,KRGKFS052} as well as z-type
CdTe~\cite{LKRSSJ06} using the VASP-PAW and FP~LMTO methods.
Although the SO splitting at the top of VB is known to play an
important role in electronic structure, and chemical bonding of
semiconductors~\cite{CW04,B90,WZ96,WCC95,CCF88,LRLSM02,KRGKFS051,KRGKFS052}
there is no systematic study of the role of the SO coupling in
optical properties of these materials.

Several attempts have been undertaken to resolve the DFT
eigenvalue problem. One such approach is utilization of the GW
approximation (``G" stands for one-particle Green's function as
derived from many-body perturbation theory and ``W" for Coulomb
screened interactions). Although GW removes most of the problems
of LDA with regard to excited state properties, it fails to
describe the semiconductors with strong Coulomb correlation
effects. For example, two studies of the band gap of ZnO
calculated using the GW correction underestimated \emph{E}$_g$ by
$1.2~\rm{eV}$\cite{UHKS02} and overestimated it by
$0.84~\rm{eV}$.\cite{OA00} Calculations for Zn, Cd, and Hg
monochalcogenides by the GW approach showed\cite{FH05} that the
band-gap underestimation is in the range 0.3--0.6~eV. Combination
of exact-exchange (EXX) DFT calculations and the
optimized-effective GW potential approach is found\cite{RQNFS05}
to improve the agreement with the experimental band gaps and
Zn-3$d$ energy levels. Band gaps calculated within the EXX
treatment are found to be in good agreement with experiment for
the $s$-$p$ semiconductors.\cite{Stadele99,SDA05} Excellent
agreement with experimental data was obtained\cite{SDA05} also for
locations of energy levels of the $d$ bands of a number of
semiconductors and insulators such as Ge, GaAs, CdS, Si, ZnS, C,
BN, Ne, Ar, Kr, and Xe.

Another means to correct the DFT eigenvalue error is to use the
screened-exchange LDA.\cite{Asahi99} Compared to LDA and GW, this
approximation is found to be computationally much less demanding,
permitting self-consistent determination of the ground-state
properties, and giving more correct band gaps and optical
properties. Other considered approaches for \textit{ab initio}
computations of optical properties involve electron--hole
interaction,\cite{BSB98} partial inclusion of dynamical vertex
corrections that neglect excitons,\cite{BTAS97} and empirical
energy-dependent self-energy correction according to the
Kohn--Sham local density theory of excitation.\cite{WK812}
However, the simplest method is to apply the scissor
operator,\cite{BS84} which displaces the LDA eigenvalues for the
unoccupied states by a rigid energy shift. Using the latter method
excellent agreement with experiments has been demonstrated for
lead monochalcogenides\cite{DREW98} and ferroelectric
NaNO$_2$.\cite{RDJEW99} However, the question as to whether the
rigid energy shift is generally applicable to semiconductors with
strong Coulomb correlation effects is open.

In this work electronic structure and optical properties of the
Zn$X$-w and -z phases have been studied in the energy range from 0
to 20~eV based on first-principles band structures calculations
derived from DFT within the LDA, GGA, and LDA+$U$.

\section{Computational details}

Experimentally determined lattice parameters have been used in the
present \textit{ab initio} calculations (Table~\ref{latparam}).
The ideal positional parameter $u$ for Zn$X$-w is calculated on
the assumption of equal nearest-neighbor bond
lengths:\cite{LYVWCC96}
\begin{equation}\label{u}
    u = \frac{1}{3} \Bigl(\frac{a}{c}\Bigr)^2 + \frac{1}{4}
\end{equation}
The values of $u$ for the ideal case agree well with the
experimental values $u^*$ (see Table~\ref{latparam}).
Self-consistent calculations were performed using a $10\times
10\times10$~mesh according to the Monkhorst--Pack scheme for the
Zn$X$-z phases, and the $\Gamma$-centered grid for the Zn$X$-w
phases.
\begingroup
\squeezetable
\begin{table*}[]
\caption{Theoretically and experimentally (in brackets) determined
unit-cell dimensions $a, c$, volumes $V$, ideal $u$ (calculated by
Eq.~\ref{u}) and experimental $u^*$ as well as values of the
parameters $U$ and $J$ from
Refs.~\onlinecite{KRGKFS051}~and~\onlinecite{KRGKFS052} were used
in the present calculations. For w-type structure $a=b$. For the
z-type structure $a=b=c$ and all atoms are in fixed positions.}
\begin{ruledtabular}
\begin{tabular}{llllllll}
\multicolumn{1}{l}{Phase} & \multicolumn{1}{c}{a, (\AA)} & \multicolumn{1}{c}{$c$~(\AA)}  & \multicolumn{1}{c}{$V$~(\AA$^3$)} & \multicolumn{1}{c}{$u^*$} & \multicolumn{1}{c}{$u$} & \multicolumn{1}{l}{$U$~(eV)} & \multicolumn{1}{l}{$J$~(eV)}\\
\hline \ \\
ZnO-w\footnotemark[1]       & 3.244(3.250) & 5.027(5.207) & $V$=45.82(47.62)  & 0.383 & 0.380 & 9 & 1 \\
ZnS-w\footnotemark[2]       & 3.854(3.811) & 6.305(6.234) & $V$=81.11(78.41)  & 0.375 & 0.375 & 6 & 1 \\
ZnSe-w\footnotemark[3]      & 4.043(3.996) & 6.703(6.626) & $V$=94.88(91.63)  & 0.375 & 0.371 & 8 & 1 \\
ZnTe-w\footnotemark[4]      & 4.366(4.320) & 7.176(7.100) & $V$=118.47(114.75)& 0.375 & 0.373 & 7 & 1 \\
\ \\
ZnO-z\footnotemark[5]       & 4.633(4.620) &              & $V$=99.45(98.61)  &       &       & 8 & 1 \\
ZnS-z\footnotemark[6]       & 5.451(5.409) &              & $V$=161.99(158.25)&       &       & 9 & 1 \\
ZnSe-z\footnotemark[7]      & 5.743(5.662) &              & $V$=189.45(181.51)&       &       & 8 & 1 \\
ZnTe-z\footnotemark[8]      & 6.187(6.101) &              & $V$=236.79(227.09)&       &       & 8 & 1 \\
\end{tabular}
\end{ruledtabular}
\label{latparam} \footnotetext[1]{Ref.~\onlinecite{ICST01}}
\footnotetext[2]{Ref.~\onlinecite{W86,XC93}}
\footnotetext[3]{Ref.~\onlinecite{ICST01,Oleg94}}
\footnotetext[4]{Ref.~\onlinecite{LRVRR03,TOM78}}
\footnotetext[5]{Ref.~\onlinecite{BD54}}
\footnotetext[6]{Ref.~\onlinecite{HM82,WLAB90}}
\footnotetext[7]{Ref.~\onlinecite{ICST01}}
\footnotetext[8]{Ref.~\onlinecite{AYA94,WLAB90}}
\end{table*}
\endgroup

\subsection{Calculations by VASP package}

Optical spectra have been studied based on the band-structure data
obtained from the VASP-PAW package,\cite{vasp} which solves the
Kohn--Sham eigenvalues in the framework of the DFT\cite{HK64}
within LDA,\cite{KS65} GGA,\cite{PBE96} and the simplified
rotationally invariant LDA+$U$.\cite{ASKCS93,DBSHS98} The exchange
and correlation energy per electron have been described by the
Perdew and Zunger parametrization\cite{PZ81} of the quantum Monte
Carlo results of Ceperley and Alder.\cite{CA80} The interaction
between electrons and atomic cores is described by means of
non-norm-conserving pseudopotentials implemented in the VASP
package.\cite{vasp} The pseudopotentials are generated in
accordance with the projector-augmented wave
(PAW)\cite{PAW94,PAW99} method. The use of the PAW
pseudopotentials addresses the problem of inadequate description
of the wave functions in the core region (common to other
pseudopotential approaches\cite{AFB01}) and its application allows
us to construct orthonormalized all-electron-like wave functions
for Zn 3$d$ and 4$s$, and $s$ and $p$ valence electrons of the $X$
atoms under consideration. LDA and GGA pseudopotentials have been
used and the completely filled semicore Zn-3$d$ shell has been
considered as valence states.

It is well known that DFT calculations within LDA and GGA locate
the Zn-3$d$ band inappropriately close to the topmost VB,
hybridizing with the O-$p$ band, falsifies the band dispersion,
and reduces the band gap. Nowadays the problem is known to be
solved by using the LDA+$SIC$ and
LDA+$U$.\cite{LRLSM02,VKP95,DPVASMACG04,JW06,KRGKFS051,KRGKFS052}
For the DFT calculations within LDA+$U$ explicit values of the
parameters $U$ and $J$ are required as input. In previous
papers\cite{KRGKFS051,KRGKFS052} we have estimated the values of
the $U$ and $J$ parameters within the constrained DFT
theory\cite{PEE98} and in a semiempirical way by performing the
calculations for different values of $U$ and forcing match to the
experimentally established\cite{RSS94} location of the Zn-3$d$
bands. Based on the results\cite{KRGKFS051,KRGKFS052} the values
of the parameters $U$ and $J$ listed in Table~\ref{latparam} are
chosen to study optical spectra.

\subsection{Calculations by MindLab package}

For investigation of the role of the SO coupling in electronic
structure and optical properties of Zn$X$,  DFT calculations have
been performed using the MindLab package,\cite{MindLab04} which
uses the full potential linear muffin-tin orbital (FP~LMTO)
method. For the core charge density, the frozen core approximation
is used. The calculations are based on LDA with the
exchange-correlation potential parametrized according to
Gunnarsson--Lundquist\cite{GL76} and
Vosko-Wilk-Nussair.\cite{VWN80} The base geometry in this
computational method consists of a muffin-tin part and an
interstitial part. The basis set is comprised of linear muffin-tin
orbitals. Inside the muffin-tin spheres the basis functions,
charge density, and potential are expanded in symmetry-adapted
spherical harmonic functions together with a radial function and a
Fourier series in the interstitial.

\subsection{Calculation of optical properties}

From the DFT calculations the imaginary part of the dielectric
function $\epsilon_2(\omega)$ has been derived by summing
transitions from occupied to unoccupied states for energies much
larger than those of the phonons:
\begin{eqnarray}
    \epsilon_2^{ij}(\omega) & = & \frac{Ve^2}{2 \pi \hbar m^2 \omega^2}
    \int d^3 {\bf k} \sum_{nn'} \langle {\bf kn} | p_i| {\bf kn'} \rangle \langle {\bf kn'}| p_j| {\bf kn} \rangle
    \times \nonumber\\
    & & f_{{\bf kn}}(1-f_{{\bf kn'}}) \delta (\epsilon_{{\bf kn'}} - \epsilon_{{\bf kn}} - \hbar \omega).
\end{eqnarray}
Here $(p_x, p_y, p_z)=\bf{p}$ is the momentum operator,
$f_{\bf{kn}}$ the Fermi distribution, and $|\bf{kn} \rangle$ the
crystal wave function corresponding to the energy
$\epsilon_{\bf{kn}}$ with momentum $\bf{k}$. Since the Zn$X$-w
phases are optically anisotropic, components of the dielectric
function corresponding to the electric field parallel ($E
\parallel c$) and perpendicular ($E \perp c$) to the
crystallographic $c$~axis have been considered. The Zn$X$-z phases
are isotropic, consequently, only one component of the dielectric
function has to be analyzed.

The real part of the dielectric function $\epsilon_1(\omega)$ is
calculated using the Kramer--Kronig transformation. The knowledge
of both the real and imaginary parts of the dielectric tensor
allows one to calculate other important optical spectra. In this
paper we present and analyze the reflectivity $R(\omega)$, the
absorption coefficient $\alpha(\omega)$, the refractive index
$n(\omega)$, and the extinction coefficient $k(\omega)$:
\begin{eqnarray}
    R(\omega) & = & \left|\frac{\sqrt{\epsilon(\omega)}-1}{\sqrt{\epsilon(\omega)}+1} \right|^2, \\
    \alpha(\omega) & = & \omega \sqrt{2
\sqrt{\epsilon_1^2(\omega)+\epsilon_2^2(\omega)}-2 \epsilon_1(\omega)}, \\
    n(\omega) & = & \sqrt{\frac{
\sqrt{\epsilon_1^2(\omega)+\epsilon_2^2(\omega)} + \epsilon_1(\omega)}{2}}, \\
k(\omega) & = & \sqrt{ \frac{
\sqrt{\epsilon_1^2(\omega)+\epsilon_2^2(\omega)} -
\epsilon_1(\omega)}{2}}. \end{eqnarray} Here
$\epsilon(\omega)=\epsilon_1(\omega)+i\epsilon_2(\omega)$ is the
complex dielectric function. The calculated optical spectra yield
unbroadened functions, and consequently have more structure than
the experimental ones.\cite{RDAJAWE97,RDJJWAE99,RDJEW99,DREW98} To
facilitate a comparison with the experimental findings, the
calculated imaginary part of the dielectric function has been
broadened. The exact form of the broadening function is unknown.
However, analysis of the available experimentally measured optical
spectra of Zn$X$ shows that the broadening usually increases with
increasing excitation energy. Also, the instrumental resolution
smears out many fine features. These features have been modelled
using the lifetime broadening technique by convoluting the
imaginary part of the dielectric function with a Lorentzian with a
full width at half maximum of 0.002$(\hbar \omega)^2$~eV and
increasing quadratically with the photon energy. The experimental
resolution was simulated by broadening the final spectra with a
Gaussian, where the FWHM is equal to 0.08 eV.

\section{Results and discussion}

\subsection{Band structure}

\begin{figure*}
\centering
\includegraphics[scale=0.75]{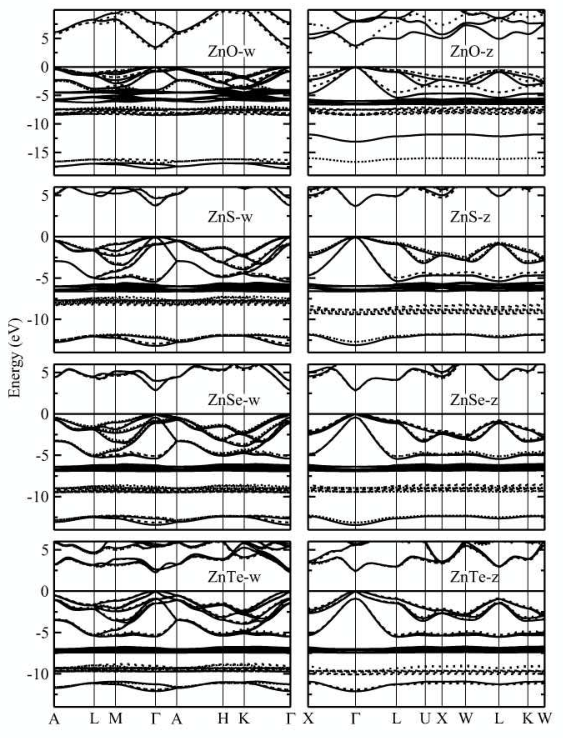}
\caption{Band dispersion for ZnO-w, ZnS-w, ZnSe-w, ZnTe-w, ZnO-z,
ZnS-z, ZnSe-z, and ZnTe-z calculated according to LDA (solid
lines) and LDA+$U$ (dotted lines). The Fermi level is set to zero
energy.} \label{dispZnY}
\end{figure*}

The optical spectra are related to band dispersion and
probabilities of interband optical transitions. So, it is of
interest to analyze the electronic structure in detail. Band
dispersions for Zn$X$-w and Zn$X$-z calculated by DFT within LDA
and LDA+$U$ are presented in Fig.~\ref{dispZnY}. The general
features of the band dispersions are in agrement with previous
studies (see, e.g., Refs.~\onlinecite{LRLSM02,
VKP95},~and~\onlinecite{VKP96}). It is seen from
Fig.~\ref{dispZnY} that the CB minimum for Zn$X$-w and Zn$X$-z are
much more dispersive than the VB maximum, which shows that the
holes are much heavier than the CB electrons in agreement with
experimental data\cite{Madelung,H73} for the effective masses and
calculated with FP~LMTO,\cite{LRLSM02} LCAO,\cite{XC93} as well as
our findings.\cite{KRGKFS051,KRGKFS051} Consequently, mobility of
electrons is higher than that of holes. Furthermore, these
features indicate that $p$ electrons of $X$ (that form the topmost
VB states) are tightly bound to their atoms, and make the VB holes
less mobile. Hence, the contribution of the holes to the
conductivity is expected to be smaller than that of CB electrons
even though the concentration of the latter is smaller than that
of the former. These features emphasize the predominant ionic
nature of the chemical bonding. Another interesting feature of the
band structures is that the VB maximum becomes more dispersive
with increasing atomic number of $X$ from O to Te.

\begin{figure}
\centering
\includegraphics[scale=0.75]{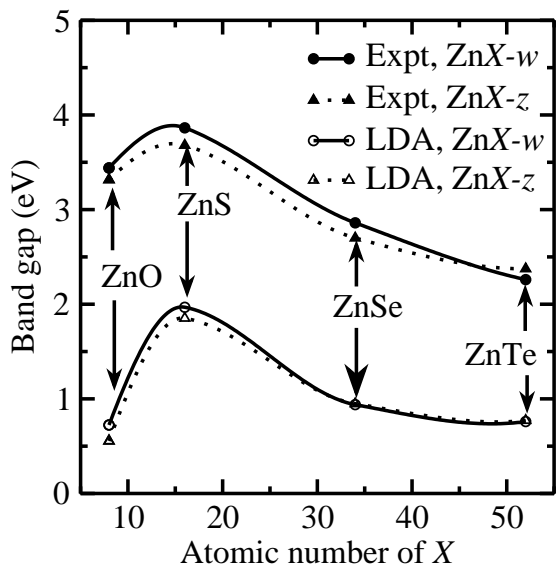}
\caption{Band gaps for Zn$X$-w (circles) and Zn$X$-z (triangles)
phases determined experimentally (filled symbols, from
Refs.~\onlinecite{KRGKFS051}~and~\onlinecite{KRGKFS052}) and
calculated (open symbols) by DFT within LDA as a function of the
atomic number of the $X$ component of Zn$X$.} \label{Eg_ZnX}
\end{figure}

As noted in our previous contributions,\cite{KRGKFS051,KRGKFS052}
the band gaps of Zn$X$ calculated by DFT within LDA, GGA, and
LDA+$U$ are underestimated and the question as to whether it is
possible to shift the CB states rigidly was kept open. As found
from the optical spectra discussed on the following sections,
rigid shifts of the CB states up to the experimentally determined
locations can provide a good first approximation for the
stipulation of the band gap. So, for the band dispersions in
Fig.~\ref{dispZnY}, we have made use of this simple way for
correcting the band gaps calculated by DFT. The only problem in
this respect was the lack of an experimental band-gap value for
ZnO-z. To solve this problem, the experimentally and calculated
(by DFT within LDA) band gaps ($E_g$) of the Zn$X$ series were
plotted as a function of the atomic number of $X$. As seen from
Fig.~\ref{Eg_ZnX}, $E_g$ for the Zn$X$-w phases are very close to
the corresponding values for the Zn$X$-z phases and the shape of
the experimental and calculated functional dependencies are in
conformity. On this basis the band gap of ZnO-z is estimated by
extrapolating the findings for Zn$X$-z from ZnS-z to ZnO-z. This
procedure gave $E_g \approx 3.3$~eV for ZnO-z.

It is well-known that not only band gaps are underestimated within
LDA and GGA, but also band dispersions come out incorrectly,
whereas location of energy levels of the Zn-3$d$ electrons are
overestimated (see, e.g.,
Refs.~\onlinecite{LC06,KRGKFS051,KRGKFS052},~and~\onlinecite{DPVASMACG04}).
As also seen from Fig.~\ref{dispZnY}, calculations within the
LDA+$U$ approach somewhat correct the location of the energy
levels of the Zn-3$d$ electrons. The elucidation of the eigenvalue
problem and the order of states at the topmost VB from LDA, GGA,
and LDA+$U$ calculations are discussed in
Refs.~\onlinecite{LC06,KRGKFS051,KRGKFS052}~and~\onlinecite{LRLSM02}
will not be repeated here.

Examination of Fig.~\ref{dispZnY} shows that the VB comprises
three regions of bands: first a lower region consists of $s$ bands
of Zn and $X$, a higher lying region where well localized Zn-3$d$
bands, and on top of this a broader band dispersion originating
from $X$-$p$ states hybridized with Zn-3$d$ states. The latter
sub-band is more pronounced in ZnO than in the other Zn$X$ phases
considered. The hybridization is most severe according to the LDA
and GGA calculations, whereas the LDA+$U$ calculations somehow
suppress this and improve the band gap underestimation. A more
detailed discussion of these aspects is found in
Refs.~\onlinecite{KRGKFS051}~and~\onlinecite{KRGKFS052}.

The SO splitting at the topmost VB is known to play an important
role for the electronic structure and chemical bonding of
solids.\cite{CW04,B90,WZ96} In semiconductors with z-type
structure the SO splitting energy is determined as the difference
between energies of the topmost VB states with symmetry
$\Gamma_{8v}$ and $\Gamma_{7v}$.\cite{CW04,B90,WZ96} In the w-type
compounds the topmost VB is split not only by SO interaction, but
also by CF giving rise to three states at the Brillouin-zone
center. To calculate the SO splitting energy for w-type phases
quasi-cubic model of Hopfield~\cite{H60} is commonly used.

It is well known that the SO splitting energy derived from
\textit{ab initio} calculations agree well with experimental data
only for some of the semiconductors. This is demonstrated, for
example, for all diamond-like group IV and z-type group III--V,
II--VI, and I--VII semiconductors,\cite{CW04} w-type AlN, GaN, and
InN,\cite{WZ96} Zn$X$-w and -z ($X$=S, Se, and
Te),\cite{KRGKFS051,KRGKFS052} and CdTe.\cite{LKRSSJ06} However,
the errors in estimated SO and CF splitting energies by LDA
calculations are significant for semiconductors with strong
Coulomb correlation effects, as demonstrated, e.g., for
ZnO.\cite{LRLSM02,KRGKFS051,KRGKFS052} For such systems DFT
calculations within LDA+$U$\cite{KRGKFS051,KRGKFS052,LRLSM02} are
shown to provide quite accurate values for $\Delta_{\rm{CF}}$ and
$\Delta_{\rm{SO}}$. Overestimation of the $p$-$d$ hybridization in
various variants of the DFT can also lead to the wrong spin-orbit
coupling of the valence bands.\cite{Shw89,Shw88}

Systematic study of the SO coupling parameters was performed for
zinc-blende II-VI semiconductors (Ref.~\onlinecite{WCC95}) using
the TB and LMTO methods, as well as for all diamondlike and
zinc-blende semiconductors (Ref.~\onlinecite{CW04}) using the
FLAPW method with and without the $p_{1/2}$ local orbitals, and
the frozen-core PAW method implemented into VASP. The corrections
coming out from inclusion of the local $p_{1/2}$ orbitals are
found to be negligible for the compounds with light atoms.
Analysis of these results shows that the So splitting energy
coming out from calculations using the VASP-PAW shows good
agreement with the experimental data. This result was also
obtained\cite{KRGKFS051} recently for Zn$X$ of wurtzite and
zinc-blende structures. As demonstrated in
Refs.~\onlinecite{KRGKFS051,KRGKFS052} the SO splitting energy
($\Delta_{\rm SO}$) increases when one moves from ZnO-z to ZnTe-z
in agreement with earlier findings of Ref.~\onlinecite{CW04}.

To study the role of the SO coupling in band dispersion the
present \textit{ab initio} calculations have been performed by
VASP and MindLab packages and spin orbit splitting energy is
found. The results are presented in Table~\ref{so}. Analysis of
the Table~\ref{so} shows that ($\Delta_{\rm SO}$) calculated by
MindLab is quite accurate.

\begingroup \squeezetable \begin{table}[h]
\caption{Calculated SO splitting energy (in meV), using the
MindLab package along with the previous theoretical and
experimental findings. }
\begin{ruledtabular}
\begin{tabular}{llll}
\multicolumn{1}{c}{ZnO-z} & \multicolumn{1}{c}{ZnS-z} & \multicolumn{1}{l}{ZnSe-z} & \multicolumn{1}{l}{ZnTe-z}\\
\hline \ \\
 31                 & 66                 & 432                 & 914                 \\
 31                 & 66                 & 432                 & 914                 \\
-34\footnotemark[1] & 66\footnotemark[1] & 393\footnotemark[1] & 889\footnotemark[1] \\
-34\footnotemark[2] & 66\footnotemark[2] & 398\footnotemark[2] & 916\footnotemark[2] \\
-37\footnotemark[3] & 64\footnotemark[3] & 392\footnotemark[3] & 898\footnotemark[3] \\
-33\footnotemark[4] & 64\footnotemark[4] & 393\footnotemark[4] & 897\footnotemark[4] \\
                    & 65\footnotemark[5] & 420\footnotemark[6] & 910\footnotemark[6] \\
\end{tabular}
\end{ruledtabular} \label{so}
\footnotetext[1]{LAPW, Ref.~\onlinecite{CW04}.}
\footnotetext[2]{LAPW+$p_{1/2}$, Ref.~\onlinecite{CW04}.}
\footnotetext[3]{VASP-PAW, Ref.~\onlinecite{CW04}.}
\footnotetext[4]{VASP-PAW, Ref.~\onlinecite{KRGKFS051}.}
\footnotetext[5]{Experiment, Ref.~\onlinecite{LB82}.}
\footnotetext[6]{Experiment, Ref.~\onlinecite{Madelung96}.}
\end{table}
\endgroup

\begin{figure*}
\centering
\includegraphics[scale=0.75]{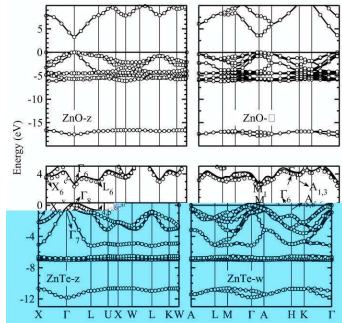}
\caption{Band dispersion for ZnO-z, ZnO-w, ZnTe-z, and ZnTe-w
calculated by the VASP-PAW method within LDA accounting for SO
coupling (solid lines) and without SO coupling (open circles).
Topmost VB of the band structure without SO coupling and center of
gravity of that with SO coupling are set at zero energy. Symmetry
labels for some of the high-symmetry points are shown for ZnTe-z
and ZnTe-w to be used for interpretation of the origin of some the
peaks in the optical spectra of Zn$X$-w and Zn$X$-z.}
\label{SOdisp}
\end{figure*}

As expected, band dispersions calculated with and without the SO
coupling differ little when the SO splitting energy is small.
However, the difference increases, when one moves from ZnO to
ZnTe. This feature is demonstrated in Table~\ref{so} and
Fig.~\ref{SOdisp} for band dispersions of ZnO-z, ZnO-w, ZnTe-z,
and ZnTe-w calculated by VASP with and without including the SO
coupling. As is well known (see e.g.
Ref.~\onlinecite{LYVWCC96,LRLSM02,KRGKFS051}), without the SO
coupling top of the VB of Zn$X$-w is split into a doublet and a
singlet states. In the band structure, Fermi level is located at
the topmost one [Fig.~\ref{SOdisp}], which is the zero energy.
Upon inclusion of the SO coupling into calculations, the doublet
and singlet states are split into three twofold degenerate states
called $A, B$, and $C$ states with energies $E_g(A), E_g(B)$ and
$E_g(C)$, respectively,\cite{MRR95} located in order of decreasing
energy, i.e. $E_g(A)>E_g(B)>E_g(C)$. The center of gravity of the
$A, B$, and $C$ states, located at $(E_g(A)-E_g(C))/3$ below the
topmost $A$ state, remains to be nearly the same as the topmost VB
corresponding to the case without the SO
coupling.\cite{LYVWCC96,LRLSM02} Consequently, to compare band
structures calculated with and without the SO coupling, one should
plot the band structure with Fermi energy at the center of gravity
of the $A, B$, and $C$ states for the former and at the topmost VB
for the latter. Hence, when the SO coupling is applied, then the
$A$ and $B$ states as well as the bottommost CB move upwards to
$(E_g(A)-E_g(C))/3$ in energy, whereas the $C$ state moves
downwards to $(E_g(A)-E_g(C))2/3$ compared to the center of
gravity. Then, positions of the lowest VB region calculated with
and without the SO coupling remains nearly identical.

\subsection{General features of optical spectra of Zn$X$}

Since optical properties of solids are based on the band
structure, the nature of the basic peaks in the optical spectra
can be interpreted in terms of the interband transitions
responsible for the peaks. Such an interpretation is available for
semiconductors with z- and w-type structures.\cite{CH65,A99,CC88}
In order to simplify the presentation of the findings of this
work, the labels $E_0, E_1$, and $E_2$ of Ref.~\onlinecite{CH65}
(from the reflectivity spectra) were retained in
Table~\ref{optical_peaks} and Fig.~\ref{SOdisp}. The subscript 0
is ascribed to transitions occurring at $\Gamma$, the subscript 1
to transitions at points in the $[111]$ direction, and the
subscript 2 to transitions at points in the $[100]$ direction
(referring to the \textbf{k} space for the z-type structure).
Assignment of the $E_0, E_1$, and $E_2$ peaks to optical
transitions at high symmetry points is presented in
Table~\ref{optical_peaks} and Fig.~\ref{SOdisp}.

\begin{figure*}[h]
\centering
\includegraphics[scale=0.9]{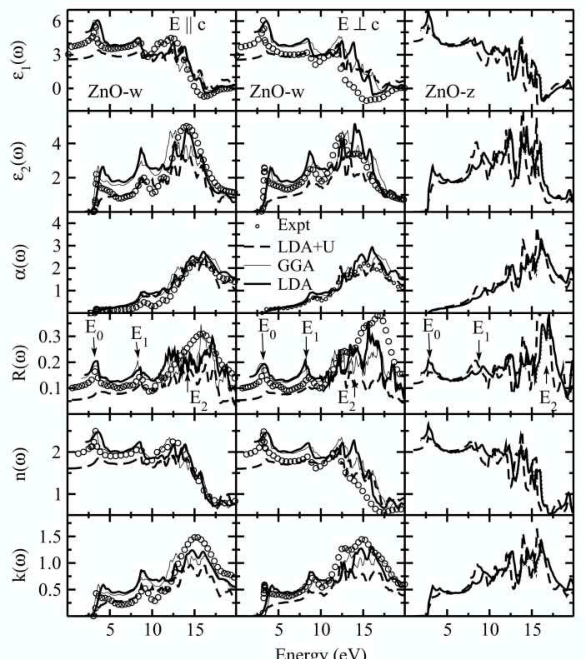} \caption{Optical
spectra of ZnO-w for $E \parallel c$ (first column) and $E \perp
c$ (second column), and ZnO-z (third column). In the first and
second columns the results obtained from calculations are plotted
by thick solid lines for LDA, thinner solid lines for GGA, and
dashed lines for LDA+$U$ as compared to experimental data from
Ref.~\onlinecite{A99} (open circles). In the third column results
calculated within LDA (solid lines), GGA (dotted lines), and
LDA+$U$ (dashed lines. $\alpha(\omega)$ is given in cm$^{-1}$
divided by 10$^5$. )} \label{ZnOwz+opt}
\end{figure*}

\begin{figure*}[h] \centering
\includegraphics[scale=0.9]{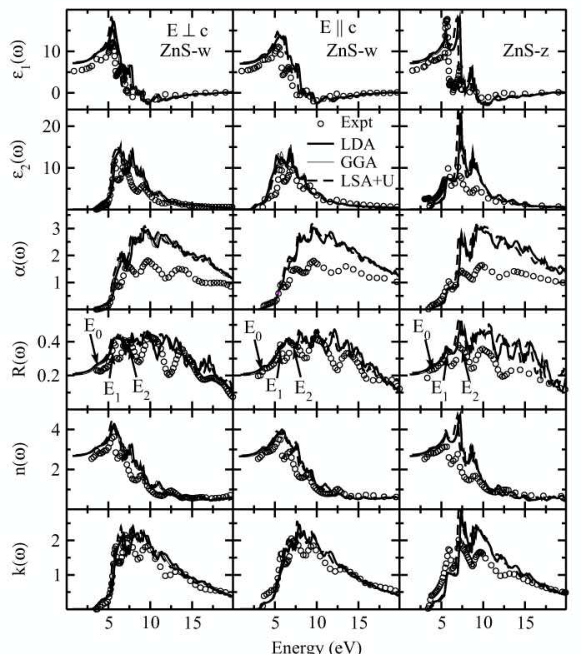} \caption{Optical
spectra of ZnS-w for $E \parallel c$ (first column) and $E \perp
c$ (second column), and ZnS-z (third column) calculated within LDA
(thick solid lines), GGA (thin solid lines), and LDA+$U$ (dashed
lines) and compared with experimental data (open circles) from
Ref.~\onlinecite{A99}. $\alpha(\omega)$ is given in cm$^{-1}$
divided by 10$^5$. } \label{ZnSwz+opt}
\end{figure*}

\begin{figure*}[h] \centering
\includegraphics[scale=0.9]{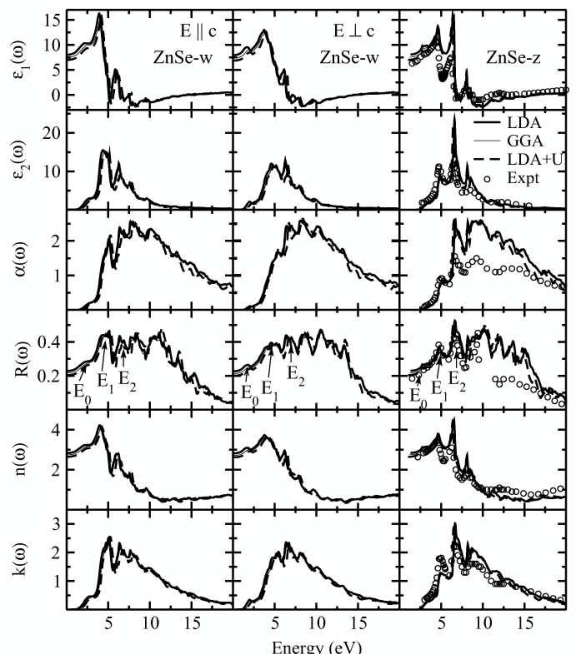}
\caption{Optical spectra of ZnSe-w for $E
\parallel c$ (first column) and $E \perp c$ (second column), and
ZnSe-z (third column) calculated within LDA (thick solid lines),
GGA (thin solid lines), and LDA+$U$ (dashed lines) and compared
with experimental data (open circles) from Ref.~\onlinecite{A99}).
$\alpha(\omega)$ is given in cm$^{-1}$ divided by 10$^5$. }
\label{ZnSewz+opt}
\end{figure*}

\begin{figure*}[h] \centering
\includegraphics[scale=0.9]{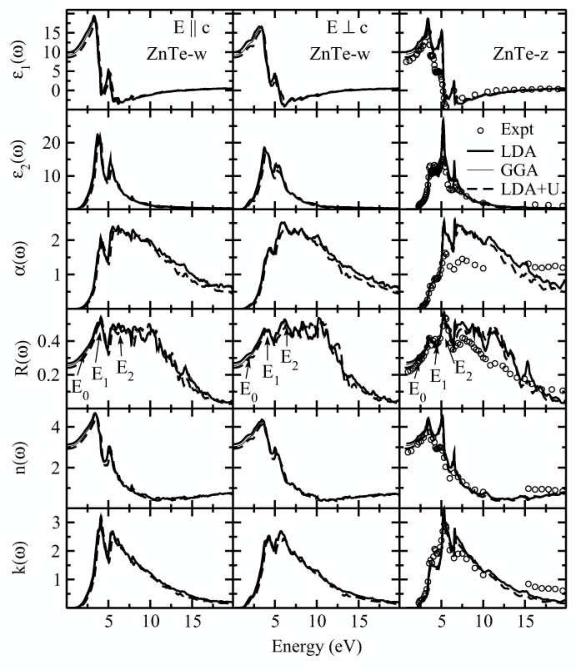}
\caption{Optical spectra of ZnTe-w for $E
\parallel c$ (first column) and $E \perp c$ (second column), and
ZnTe-z (third column) calculated within LDA (thick solid lines),
GGA (thin solid lines) and LDA+$U$ (dashed lines) and compared
with experimental data (open circles) from Ref.~\onlinecite{A99}.
$\alpha(\omega)$ is given in cm$^{-1}$ divided by 10$^5$. }
\label{ZnTewz+opt}
\end{figure*}

The optical spectra $\epsilon_1(\omega), \epsilon_2(\omega),
\alpha(\omega), R(\omega), n(\omega)$, and $k(\omega)$ calculated
by DFT within LDA, GGA, and LDA+$U$ are displayed in
Figs.~\ref{ZnOwz+opt}--\ref{ZnTewz+opt} and compared with
available experimental findings.\cite{A99} The spectral profiles
are indeed very similar to each other. Therefore, we shall only
give a brief account, mainly focusing on the location of the
interband optical transitions. The peak structures in
Figs.~\ref{ZnOwz+opt}--\ref{ZnTewz+opt} can be explained from the
band structure discussed above.

\begingroup
\squeezetable
\begin{table}[h]
\caption{Relation of the basic $E_0, E_1$, and $E_2$ peaks in the
optical spectra of Zn$X$ to high-symmetry points (see
Refs.~\onlinecite{CH65}~and~\onlinecite{A99}) in the Brillouin
zone at which the transitions seem to occur.}
\begin{ruledtabular}
\begin{tabular}{llll}
\multicolumn{1}{l}{Peak}  & \multicolumn{1}{c}{z-type} & \multicolumn{1}{c}{w-type, $E \parallel c$}  & \multicolumn{1}{c}{w-type, $E \perp c$} \\
\hline \ \\
$E_0$ & $\Gamma_8 \rightarrow \Gamma_6$ & $\Gamma_1 \rightarrow \Gamma_1$ & $\Gamma_6 \rightarrow \Gamma_1$ \\
$E_1$ & $L_{4,5} \rightarrow L_6$       & $A_{5,6} \rightarrow A_{1,3} $  & $M_4 \rightarrow M_1$           \\
$E_2$ & $X_7 \rightarrow X_6$           &                                 &                                 \\
\end{tabular}
\end{ruledtabular} \label{optical_peaks}
\end{table}
\endgroup

All peaks observed by experiments (see, e.g.,
Refs.~\onlinecite{CH65} and~\onlinecite{A99}) are reproduced by
the theoretical calculations. Because of the underestimation of
the optical band gaps in the DFT calculations the locations of all
the peaks in the spectral profiles are consistently shifted toward
lower energies as compared with the experimentally determined
spectra. Rigid shift (by the scissor operator) of the optical
spectra has been applied, which somewhat removed the discrepancy
between the theoretical and experimental results. In general, the
calculated optical spectra qualitatively agree with the
experimental data. In our theoretical calculations the intensity
of the major peaks are underestimated, while the intensity of some
of the shoulders are overestimated. This result is in good
agreement with previous theoretical findings (see, e.g.,
Ref.~\onlinecite{WK812}). The discrepancies are probably
originating from the neglect of the Coulomb interaction between
free electrons and holes (excitons), overestimation of the optical
matrix elements, and local-field and finite-lifetime effects.
Furthermore, for calculations of the imaginary part of the
dielectric response function, only the optical transitions from
occupied to unoccupied states with fixed $\mathbf{k}$ vector are
considered. Moreover, the experimental resolution smears out many
fine features and, as demonstrated in Fig.~\ref{exp_ZnO}, there is
inconsistency between the experimental data measured by the same
method and temperature. However, as noted in Introduction, by
accounting for the excitons and Coulomb correlation effects in
\textit{ab initio} calculations\cite{LC06} by the
linearized-augmented plane-wave method plus local orbitals
(LAPW+LO) within LDA+$U$ not only corrected energy position of the
Zn-3$d$ electrons and eigenvalues, but also optical response.
Consequently, accounts for the excitons play an important role in
the optical spectra.

The optical spectra calculated within LDA, GGA, and LDA+$U$ do not
differ significantly from each other for the Zn$X$-w and -z phases
except for ZnO-w and -z, for which the optical spectra calculated
within LDA+$U$ are significantly different from those obtained by
LDA and GGA. The difference between the optical spectra calculated
by LDA and GGA and those calculated by LDA+$U$ decreases, when one
moves from ZnO to ZnTe. For ZnTe the difference can be said to be
very small. This feature shows that in ZnO-w and -z Coulomb
correlation effects are strong compared to the other Zn$X$-w and
-z phases in agreement with recent LAPW+LO calculations\cite{LC06}
including electron--hole correlations.

Comparison of the optical spectra for $E \perp c$ and $E
\parallel c$ for each of the Zn$X$-w phases with the isotropic spectra of
the corresponding Zn$X$-z phases shows that the locations of the
peaks almost coincide. This similarity reflects that there is only
small differences in the local arrangement of the atoms in the
Zn$X$-w and corresponding -z phases.

\subsection{ZnO-w and ZnO-z}
The optical spectra of ZnO-w and -z calculated by DFT within LDA,
GGA and LDA+$U$, together with measured data are displayed in
Fig.~\ref{ZnOwz+opt}. One clearly sees three major peaks in the
experimental spectra located in the energy ranges 3.1--3.3
($E_0$), 7.5--8.5 ($E_1$) and 10--15~eV ($E_2$). In the $E_2$ peak
$\epsilon_2(\omega), \alpha(\omega), R(\omega)$, and $k(\omega)$
are seen to take larger values than those in the $E_0$ and $E_1$
peaks. This is one of the major features, which distinguishes
ZnO-w and -z from the other Zn$X$ phases.

It should be noted that with increasing value for the parameter
$U$ in the LDA+$U$ calculations, the intensity of the $E_0$ and
$E_1$ peaks of ZnO-w decreases compared with the LDA and GGA
findings as well as with the experimental data. However, the
intensity of the $E_1$ peak of ZnO-z from the LDA+$U$ calculations
has increased and has become even larger than those derived from
the LDA and GGA calculations as well as the experimental data. The
intensity of $E_2$ peak from the LDA+$U$ calculations oscillates
significantly, showing disagreement with the LDA and GGA
calculations as well as the experimental measurements. Hence,
although LDA+$U$ calculations\cite{KRGKFS051} were good to
increase the LDA-derived band gap and the SO splitting energy as
well as to decrease the crystal-field splitting energy and
improving the band dispersion, it was poorer than LDA and GGA to
describe the optical properties of the ZnO phases. Probably, this
discrepancy comes about because in our \textit{ab initio}
calculations electron--hole interactions and SO coupling are not
included.\cite{LC06} The strong variation of the optical
properties with increasing the values of $U$ indicates appreciable
Coulomb correlation effects in ZnO-w and -z in agreement with our
previous band structure findings\cite{KRGKFS051,KRGKFS052} and
LAPW+LO calculations\cite{LC06} including excitonic effect. This
feature is not present in the spectra for the other Zn$X$-w and -z
phases considered.

For convenience of analysis the $\epsilon_2(\omega)$ profile was
analyzed by adjusting the peak location to the experimental data
of Ref.~\onlinecite{A99} by rigid shift. On comparing this result
with that of Ref.~\onlinecite{KNPSW71}, it is concluded that the
peaks at 3.40~eV for $E \parallel c$ and that at 3.33~eV for $E
\perp c$ of $\epsilon_2(\omega)$ and $R(\omega)$ can be ascribed
to transitions at the fundamental absorption edge. As shown in
Ref.~\onlinecite{KNPSW71}, the energy difference (0.07~eV) between
these two peaks gives the separation between the so-called
\textit{A}, \textit{B} (for $E \perp c$), and \textit{C} (for $E
\parallel c$)  states forming the topmost VB of w-type semiconductors
in agreement with 0.083~eV according to the band-structure
analyses in
Refs.~\onlinecite{KRGKFS051}~and~\onlinecite{KRGKFS052}.

There are two broad shoulders of the peak $E_0$ located at 4.44
and 5.90~eV for $E \perp c$, and 3.90 and 5.29~eV for $E
\parallel c$. Similar shoulders are found at lower energies in the experimental
spectra of Refs.~\onlinecite{KNPSW71}~and~\onlinecite{LY68}
observed at 3.35 and 3.41~eV for $E \perp c$ and 3.39 and 3.45~eV
for $E \parallel c$, and the origin of these shoulders have been
ascribed to exciton--phonon coupling. However, in our \textit{ab
initio} studies excitons and lattice vibrations are not taken into
consideration.

\subsection{ZnS-w and ZnS-z}
The experimental\cite{A99} optical spectra for the ZnS-w and -z
phases are displayed in Fig.~\ref{ZnSwz+opt} together with those
calculated according to the LDA, GGA, and LDA+$U$. It is seen that
the magnitudes of the experimentally\cite{CH65} observed shoulders
around the $E_2$ peak in the reflectivity spectra of ZnS-w are
overestimated in the DFT calculations. As a result, the
intensities of the shoulders are almost the same as intensities of
the peaks $E_1$ and $E_2$ for $E \perp c$ and even exceeds them
for $E \parallel c$.

The calculated optical spectra for ZnS-w by LDA, GGA, and LDA+$U$
turned out to be almost identical at energies below 10~eV.
However, at higher energies the LDA- and GGA-derived peaks differ
from those obtained by LDA+$U$. This difference can be associated
with Zn-3$d$ electrons, which were shifted toward lower energies
in the LDA+$U$ calculations. Hence, in ZnS-w and -z the Coulomb
correlation effects appear to play a significant role in optical
properties at energies higher than 10~eV.

Compared to ZnO-w, the calculated optical spectra of ZnS-w and -z
show larger disagreement with the experimental data. The
discrepancy is quite pronounced in the absorption and reflectivity
spectra of ZnS-w and -z, especially at energies exceeding 7~eV.
The magnitude of the peaks located at higher energies are
overestimated significantly compared to the experimental data. The
overestimation is more severe in ZnS-z than in ZnS-w as judged
from the intensity of the $E_2$ peak.

\subsection{ZnSe-w, ZnTe-w, ZnSe-z, and ZnSe-z}
The optical spectra for ZnSe-w, ZnSe-z, ZnTe-w, and ZnTe-z
calculated by DFT within LDA, GGA, and LDA+$U$ are displayed in
Figs.~\ref{ZnSewz+opt}~and~\ref{ZnTewz+opt} together with
corresponding experimental spectra. Since experimental optical
spectra for ZnSe-w and ZnTe-w are not available, rigid shift of
the parameters toward higher energies have been performed on the
basis of the reflectivity spectra for ZnSe-z and ZnTe-z
(Figs.~\ref{ZnSewz+opt}~and~\ref{ZnTewz+opt}). A closer inspection
of Figs.~\ref{ZnSewz+opt}~and~\ref{ZnTewz+opt} shows that the
optical spectra calculated within LDA, GGA, and LDA+$U$ are almost
the same for all selenide and telluride phases. The small
differences noted in the absorption and reflectivity spectra
appear to originate from the Zn-3$d$ electrons.

The location and magnitude of the experimentally measured $E_1$
peak in the reflectivity spectra of ZnSe-z and ZnTe-z have been
assigned\cite{A99,CH65}  to fundamental absorption and $\Lambda_3
- \Lambda_1$ transitions at the $[0.17, 0.17, 0.17]$ point of the
Brillouin zone. These assignments agree well with theoretical
calculations. However, the theoretical calculations did not locate
the $E_1 + \Delta_1$ peak on the high-energy side of the $E_1$
peak, which was observed experimentally for both ZnSe-z and
ZnTe-z. The reason is certainly that SO coupling was not included
in the calculations.

The experimental~\cite{CG63} $E_0$ peak in the reflectivity
spectra, corresponding to transitions at ${\bf k=0}$ [viz. from
the highest state of VB ($\Gamma_{15}$) to the lowest state of CB
($\Gamma_1$)] is well reproduced by the theoretical calculations.
One also sees the $E_0+\Delta_0$ peak in the theoretical spectra,
which previously\cite{CG63} was ascribed to SO splitting. Since SO
coupling was neglected in the theoretical calculations, the origin
of the $E_0+\Delta_0$ peak is not likely to be related to SO
coupling.

Similar to the findings for ZnS-z, theoretically calculated
optical spectra for the lower energy regions of ZnSe-z and ZnTe-z
agree with experimental findings. However, the intensity of the
peaks located at higher energies are overestimated in the DFT
calculations. Anyway, this discrepancy is not as severe as that
for ZnS-z. The calculated reflectivity spectra agree well with
experimental data in the energy range $\leq 6$~eV. At higher
energies (6--15~eV) LDA, GGA, and LDA+$U$ all overestimate the
intensity of the reflectivity. Fairly good agreement with the
experimental data is achieved in the energy range 15--20~eV for
the real and imaginary parts of the dielectric function for ZnSe-z
and ZnTe-z. For the other optical spectra of ZnSe-z and ZnTe-z the
agreement between theory and experiment is poorer.

\subsection{Influence of spin-orbit splitting on the optical spectra of Zn$X$}

\begin{figure}[h] \centering
\includegraphics[scale=0.75]{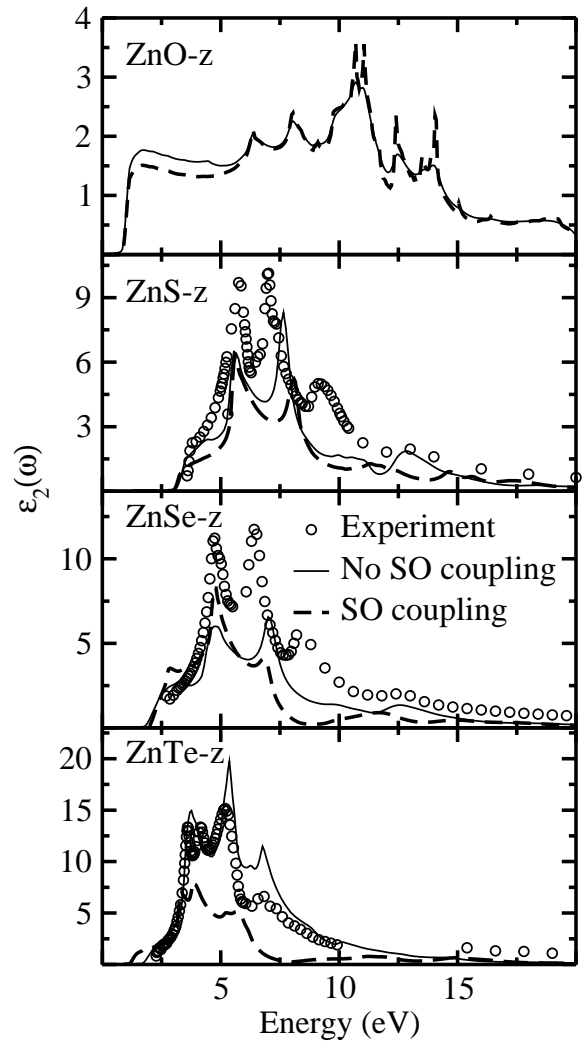}
\caption{Imaginary part of the dielectric response function
$\epsilon_2(\omega)$ for ZnO-z, ZnS-z, ZnSe-z, and ZnTe-z
calculated by DFT using the MindLab package within LDA including
SO coupling into consideration (dashed lines) and without SO
coupling (continuous lines) and compared to experimental data
(open circles) from Ref.~\onlinecite{A99}. } \label{zSO+optics}
\end{figure}

It is well-known that SO splitting at the top of the VB of a
semiconductor is very important for optical transitions and one
should expect large difference in the optical spectra calculated
with and without the SO coupling. In this section we shall analyze
how the SO coupling influences the optical spectra of Zn$X$. For
this analysis, \textit{ab initio} band structure calculations have
been performed using the MindLab software with and without SO
coupling. Based on the band structure studies, dielectric response
function $\epsilon_2(\omega)$ has been calculated. The results for
Zn$X$-z are presented in Fig.~\ref{zSO+optics} and compared with
experimental data, where it is seen that $\epsilon_2(\omega)$
calculated for ZnO-z without SO coupling is slightly higher than
that with the SO coupling, the main deviations occurring at
3.44--6.00~eV and 10.00--12.00~eV. The reason for the small
distinctions in $\epsilon_2(\omega)$ in this case is the small SO
splitting energy.

Our findings show that the SO splitting influence the calculated
optical spectra and in particular it most pronounced at energies
lower than $\sim 12$~eV. At higher energies, the difference
between the optical spectra calculated with and without SO
coupling is fairly small and agree reasonable well with the
experimental data.\cite{A99} This statement applies to all Zn$X$
phases studied.  It should be noted that intensities of the peaks
calculated with the SO coupling are generally lower than those
obtained without SO coupling and the latter set agrees better with
experimental data than the former. Furthermore, in the
experimental spectra there are low intensity peaks located at
9.4~eV in ZnS-z, 8.4~eV in ZnSe-z, and 7.0~eV in ZnTe-z. However,
these peaks are not seen in the calculated spectra with the SO
coupling. As noted, this discrepancy can be related to neglect of
many above factors like Coulomb interaction between electrons and
holes, local-field effects, and indirect transitions etc.

\subsection{Role of the ground state structure in the optical spectra of Zn$X$}

\begin{figure}[h] \centering
\includegraphics[scale=0.75]{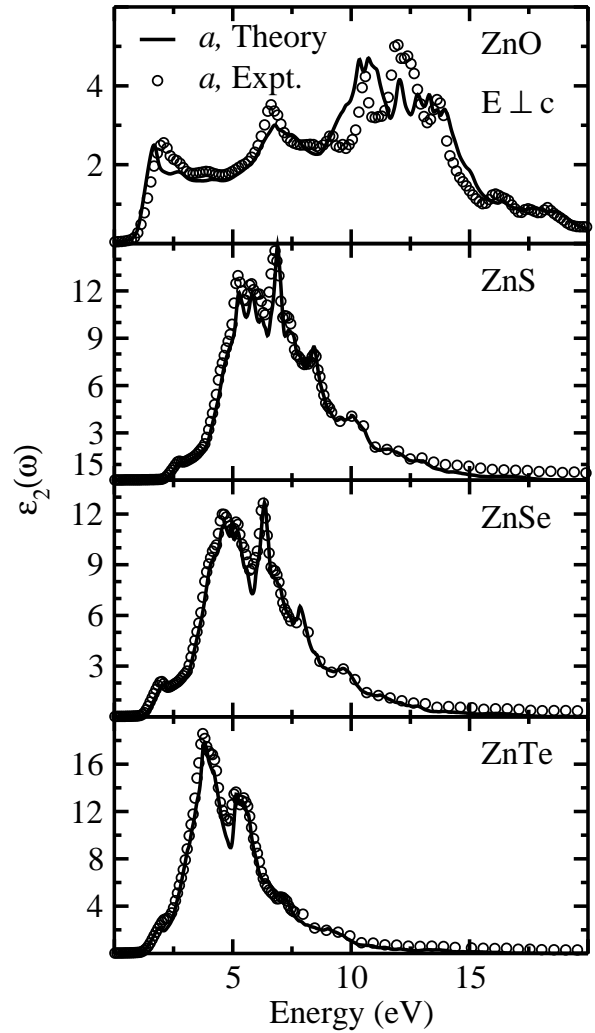}
\caption{Imaginary part of the dielectric response function
$\epsilon_2(\omega)$ for Zn$X$-z calculated by VASP-PAW package
using the theoretically (solid lines) and experimentally (open
circles) determined lattice parameters. } \label{xy}
\end{figure}

In this section we shall analyze optical spectra of Zn$X$
calculated using the experimentally and theoretically determined
lattice parameters. To find the lattice parameters from the $ab
initio$ calculations, the structural optimization has been
performed, which includes the following steps: (i) atoms are
relaxed keeping the volume and shape of the lattice. After
convergence is reached, the resulting lattice and positional
parameters have been used as input to optimize atomic positions,
shape and volume of the unit cell altogether. Then dependence of
the total energy on volume is studied. Minimum of the dependence
was accepted as the equilibrium state. Lattice and positional
parameters corresponding to the minimum is referred to as the
theoretically determined lattice parameters. The thus determined
theoretical lattice parameters do not deviate much from
experimental ones. The parameters along with the experimentally
determined ones were used for subsequent computations of the
electronic structure and optical spectra. The results are
presented in Fig.~\ref{xy} for Zn$X$-w for $E \perp $. Analysis
shows that the optical spectra of ZnO-w for $E \perp $ deviate
each from other at lower energies corresponding to the fundamental
absorption and at higher energies in the range ~8--13~eV. The
reason of the difference can be related to changes of the $p$-$d$
coupling because of the changes of the Zn-O bond lengths coming
out from structural optimization. Optical spectra of ZnO-w for $E
\parallel c$ and those of other Zn$X$-w and -z calculated using
the theoretical and experimental lattice parameters do not differ
each from other significantly.

\section{Conclusion}
The band structures of the Zn$X$-w and -z phases ($X$=O, S, Se,
and Te) are calculated by DFT within LDA, GGA, and LDA+$U$. The
topmost VB states are found to be more dispersive than the
bottommost CB states. Spin-orbit coupling is found to play an
important role for band dispersion, location and width of the
Zn-3$d$ band, and the lowest $s$ band. By analyzing the dependence
of the band gaps on the atomic number of $X$ for Zn$X$, the band
gap of ZnO-z is estimated to be $\sim3.3$~eV. Using the electronic
band structures as references, the optical spectra of Zn$X$-w and
-z are analyzed in the energy range 0--20~eV. The locations of the
peaks corresponding to transitions at the fundamental absorption
edge calculated by DFT are shifted to lower energies relative to
the experimental peaks. This deficiency originates from the
well-known errors in band gaps calculated according to DFT. In
order to correct the underestimation of band gaps calculated by
DFT, the location of the calculated peaks of the optical spectra
have been rigidly shifted toward higher energies to match with the
experimentally determined locations. In the thus obtained spectra
the locations of the peaks in the lower energy region agree well
with the experimental data. However, the peaks in the higher
energy region agree only tolerably well with the experimental
findings. The overall conclusion is that the ${\bf k}$-independent
scissors operator provides a good first approximation for
correlation of the underestimated band gaps for the Zn$X$-w and -z
phases. Based on this result, ``corrected" band structures of the
Zn$X$ phases are arranged by adjusting the band gap up to
experimentally measured value (viz. rigidly lifting the lowest
CB). Not only the locations, but also the intensities of some of
the calculated low-energy peaks agree with available experimental
data for all Zn$X$ phases. However, the intensities of some peaks
located at higher energies and shoulders have been overestimated.
The GGA approach slightly improved the band-gap values. Also, the
optical spectra of ZnO-w for $E \perp c$ calculated within the GGA
agree better with the experimental data than those calculated
within the LDA and LDA+$U$ approaches. The value for the
corresponding transition at the fundamental absorption edge is
decreased and becomes sharper with the use of the GGA thus
providing better agreement with experimental data than LDA. For $E
\perp c$ inhomogeneity in the electron gas plays an important
role, while it is not so important for $E \parallel c$. The
optical spectra for ZnO-w and -z calculated within LDA+$U$ for the
energy range 0--20~eV is found to depend significantly on the
location of energy levels of the Zn-3$d$ electrons. For the other
Zn$X$-w and -z phases such changes are not so pronounced, in fact,
only noticeable at energies above 10~eV. Strong Coulomb
correlation effects are established for ZnO-w and -z. According to
the present LDA+$U$ calculations the probability for the optical
transitions at the fundamental absorption edge of ZnO-w and -z
decrease with increasing $U$. Optical spectra for ZnO-z, ZnSe-w,
and ZnTe-w have been predicted. The influence of the spin-orbit
coupling is found to increase with increasing the atomic number of
the $X$ component of Zn$X$.

\section*{Acknowledgments}
This work has received financial and supercomputing support from
the Research Council of Norway and Academy of Sciences of
Uzbekistan (Project N31-36,24-06). The authors are thankful to R.
Vidya for critical reading of the manuscript and comments. Also we
thank Dr. P. Vajeeston, A. Klaveness, and Dr. K. Knizek for
computation-practical help.


\end{document}